\documentclass[12pt,preprint]{aastex}

\usepackage{epsfig}
\usepackage{CJK}

\shorttitle{Fitting of spectra using PAH mixtures}
\shortauthors{Zhang and Kwok}

\begin{document}

\title{On the viability of the PAH model as an explanation of the unidentified infrared emission features}

\begin{CJK*}{Bg5}{bsmi}
\CJKtilde
\author{Yong Zhang (±iªa) and Sun Kwok (³¢·s)}
\affil{Department of Physics, The University of Hong Kong, Pokfulam, Hong Kong, China}
\email{zhangy96@hku.hk; sunkwok@hku.hk}

\begin{abstract}

Polycyclic aromatic hydrocarbon (PAH) molecules are widely considered as the preferred candidate for the carrier of the unidentified infrared emission bands observed in the interstellar medium and circumstellar envelopes.  In this paper we report the result of fitting a variety of non-PAH spectra (silicates, hydrogenated amorphous carbon, coal and even artificial spectra) using the theoretical infrared spectra of PAHs from the NASA Ames PAH IR Spectroscopic Database. We show that these non-PAH spectra can be well fitted by PAH mixtures. This suggest that a general match between astronomical spectra and those of PAH mixtures does not necessarily provide definitive support for the PAH hypothesis. 

\end{abstract}


\section{Introduction}

The unidentified infrared emission (UIE) bands
at 3.3, 6.2, 7.7--7.9, 8.6, 11.3 and 12.7\,$\mu$m have been detected
in a variety of astronomical objects 
\citep[for a recent review, see][]{pe13}. These UIE bands have long been known to arise from C$-$H and C$-$C vibrational modes of aromatic compounds \citep{dw81} and are therefore also referred to as aromatic infrared bands (AIBs).  A wide variety of hydrocarbon and carbonaceous materials comprising aromatic units have been proposed as possible carriers of the UIE bands.  These include hydrogenated amorphous carbon \citep[HAC,][]{duley1983}, quenched carbon composites \citep[QCC,][]{sak87}, polycyclic aromatic hydrocarbon (PAH) molecules \citep{lp84,al85,al89,pl89}, coal \citep{papoular1989}, and petroleum fractions \citep{cataldo2003}.
Among these models, the PAH hypothesis has been by far the most popular, and the UIE bands are frequently referred to in the literature simply as PAH bands. The case for the PAH hypothesis has been summarized in \citet{tie08}.

In spite of the popularity of the PAH hypothesis, there has been increasing doubts about whether PAH molecules can adequately explain the UIE phenomenon.  
The UIE bands in astronomical spectra are usually accompanied by aliphatic bands at 3.4, 
6.9 and 7.3\,$\mu$m and unassigned features at 15.8, 16.4, 17.4, 17.8 and 
18.9\,$\mu$m \citep{jd90, kv99, ct00, sturm2000, sellgren2007}.  In order to account for the aliphatic bands,  the PAH model has  been modified to include a small number of methyl side groups \citep{ld12,yg13}.
Furthermore, the UIE bands are often superimposed upon strong broad emission plateaus at 6--9 and 
10--15\,$\mu$m that have been attributed to bending modes emitted by a mixture of alkane 
and alkene groups \citep{kv01}, implying a significant contribution from the aliphatic component.
These observational properties motivate us to invoke an alternative model, mixed aromatic/aliphatic organic 
nanoparticles (MAONs), as a possible alternative carrier of UIE bands \citep{kz11,kz13}. The disorganized molecular 
structure of MAONs, i.e. aromatic rings ($sp^2$ bonded) linked by
aliphatic chains ($sp^3$ bonded), are fundamentally different from the chemical structures of PAHs, which are two-dimensional and regular in structure.  MAONs represent an extension of the family of amorphous hydrocarbons that occur naturally in nature and artificially produced in the laboratory.    
Amorphous hydrocarbons are often the natural product of combustion and it is conceivable that similar products can be produced in the circumstellar environment \citep{col1997}.  

The general agreement between the overall appearance of astronomical spectra and that of PAH mixtures 
has been widely taken as evidence for the success of the PAH model \citep[see e.g., Figure 11,][]{tie08}.  \citet{ah99} and \citet{ca11} have reproduced the 5--15 $\mu$m spectra  of astronomical objects using experimental and theoretical spectra of PAHs from the NASA Ames PAH IR Spectroscopic Database  \citep[PAHdb;][]{bb10,bb14}.   Such a spectral synthesis requires a large variety of PAHs with different sizes, structures, and charge states, each of which has arbitrary abundance.  With a large number of free parameters, such a decomposition suffers from  non-uniqueness and cannot be used to identify individual PAH molecules.   A more useful exercise  is to divide these PAH spectra into a few subclasses according to their sizes, charge states, or compositions and investigate the variations of these PAH populations in different environments using the PAHdb fitting \citep{rb11,rb14}.  \citet{bb13} found that the template spectra for the size and charge subclasses created from the PAHdb can be shown to correlate with those derived from astronomical spectra derived from blind signal separation (BSS) methods \citep{ber07,pil12}. Specifically, the spatially resolved spectra of the reflection nebula NGC\,7023  have been fitted with these templates.    

In order to test the validity of the PAH-fitting exercise, we have used the same database and model to fit several spectra obtained from structures which are not made of PAHs.  The purpose of the exercise is to test whether  the PAHdb model can  fit the non-PAH spectra equally well.  If it can, then it implies that the fact that the model can fit the astronomical UIE spectra is not evidence for PAH mixtures as the carrier of UIE.

\section{Test of the PAHdb model}

For the spectral fitting, we made use of Version 2.00 of the PAHdb and the IDL package {\it AmesPAHdbIDLSuite} developed by \citet{bb14} \footnote{http://www.astrochem.org/pahdb}. The database contains 700 computational and 75 experimental spectra of PAH molecules and ions.  The sizes of the PAH molecules range from 6 to 384 carbon atoms and the charged states include neutral, anion ($-$), and cations ($+, ++$ and $+++$).  This is the primary database and tool that have been used to fit astronomical UIE spectra in support of the PAH model \citep[see e.g.,][]{ca11, bb13}.  

A non-negative least-squares algorithm was applied to synthesize the target spectra from  a total of 700 theoretical PAH spectra in PAHdb. We assume that all PAHs thermally emit at a temperature of 500\,K.  In order to be consistent with previous fittings using the PAHdb model, we have convolved the PAH spectra by a Lorentzian profile with a fixed width of 
15\,cm$^{-1}$, and redshift all features by 15\,cm$^{-1}$ to compensate the anharmonic effect \citep{rb14}.
  
In order to quantitatively evaluate the goodness of the fittings, we computed the $\chi^2/dof$ for each of the fittings, where $dof$ is the degree of freedom, defined as the number of wavelength bins of the observed spectra.
The values of $\chi^2/dof$ in each of the fittings are given in the respective figure captions.


The PAHdb fitting was performed to several non-PAH spectra, including the laboratory spectra of HAC and coal, the theoretical spectrum of a $sp^2$/$sp^3$ bonded molecule C$_{55}$H$_{56}$, astronomical spectrum of amorphous silicates, and a number of artificially generated spectra with random wavelengths and intensities.  In general, neutral, anionic and cationic PAHs are all needed in the fitting.  Typically, the models include about 40 different molecular species. In all the cases, the contributions from small PAHs ($N_c <$ 50) are more significant than large PAHs. This is similar to the fittings of UIE spectra in previous works \citep{bb14}. The relative contributions from different charge states change from case to case.

\begin{figure*}
\epsfig{file=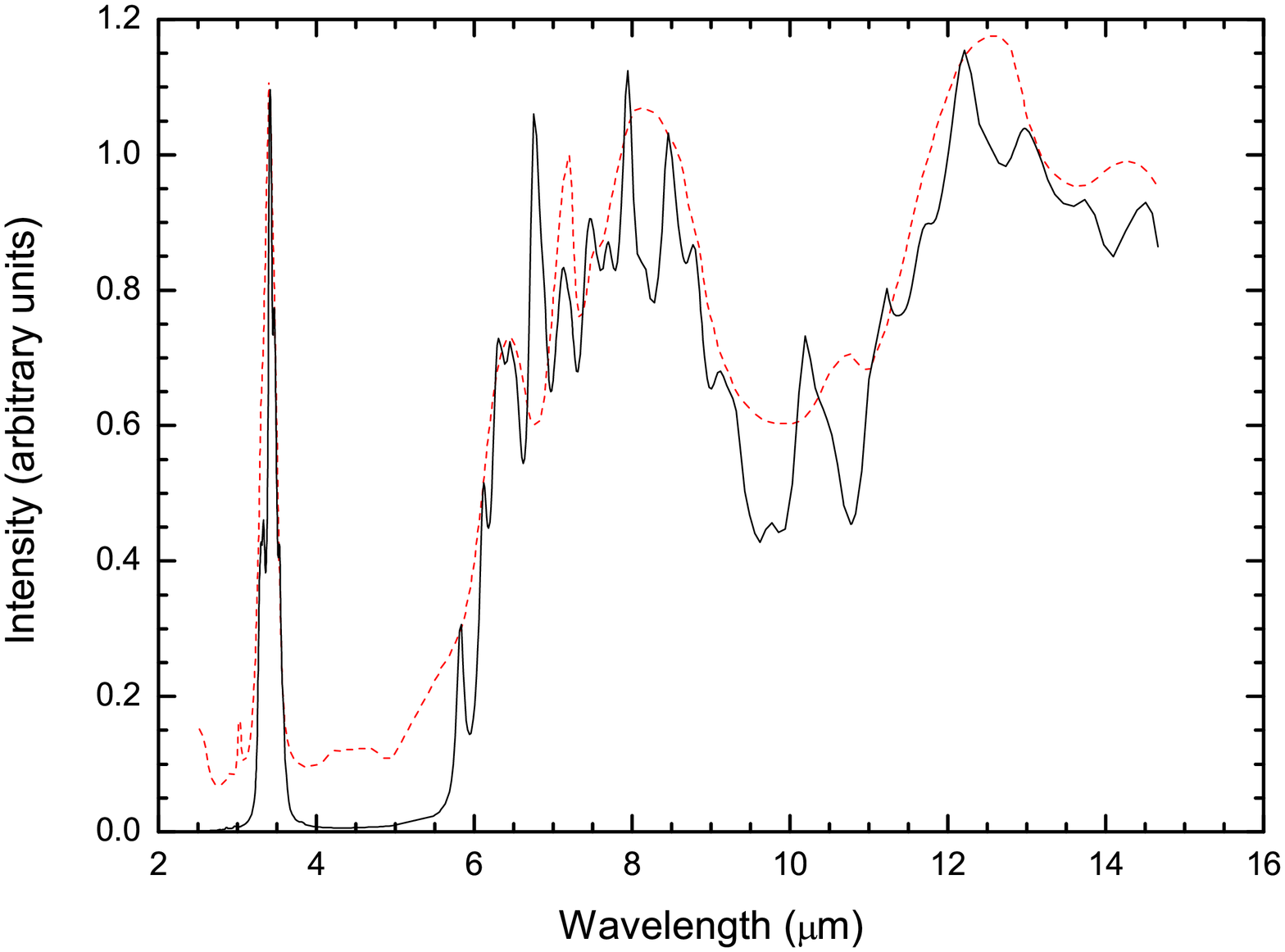, height=12cm}
\caption{Fitting of the laboratory infrared spectrum (solid line) of HAC obtained by \citet{dis83} using the PAHdb model (dashed line). The value of $\chi^2/dof$ is 0.64.
}
\label{Dis}
\end{figure*}

The HAC spectrum taken from \citet{dis83}  shows two strong plateau features at 6--9 and 10--15\,$\mu$m (Figure~\ref{Dis}). Such plateau features are commonly found in the spectra of aliphatic-rich soot samples  \citep{pi08}, in spectra of coal \citep{guillois1996}, and in spectra of astronomical sources showing strong aliphatic bands \citep{kv01}. 
Although the spectra of individual PAH molecules do not show such broad bands, 
Figure~\ref{Dis} shows that the spectrum of the HAC materials obtained in laboratory can be well fitted using the PAHdb model. Given the large number and diverseness of the PAH molecules used in the PAHdb model, emission features can be found at almost any wavelength within the 5--15\,$\mu$m range. 
The emission plateaus can easily be artificially reproduced by blending many PAH bands together.
Therefore, the fitting of infrared emission using PAH spectra may not necessarily imply that PAH molecules are the emitter.

\begin{figure*}
\epsfig{file=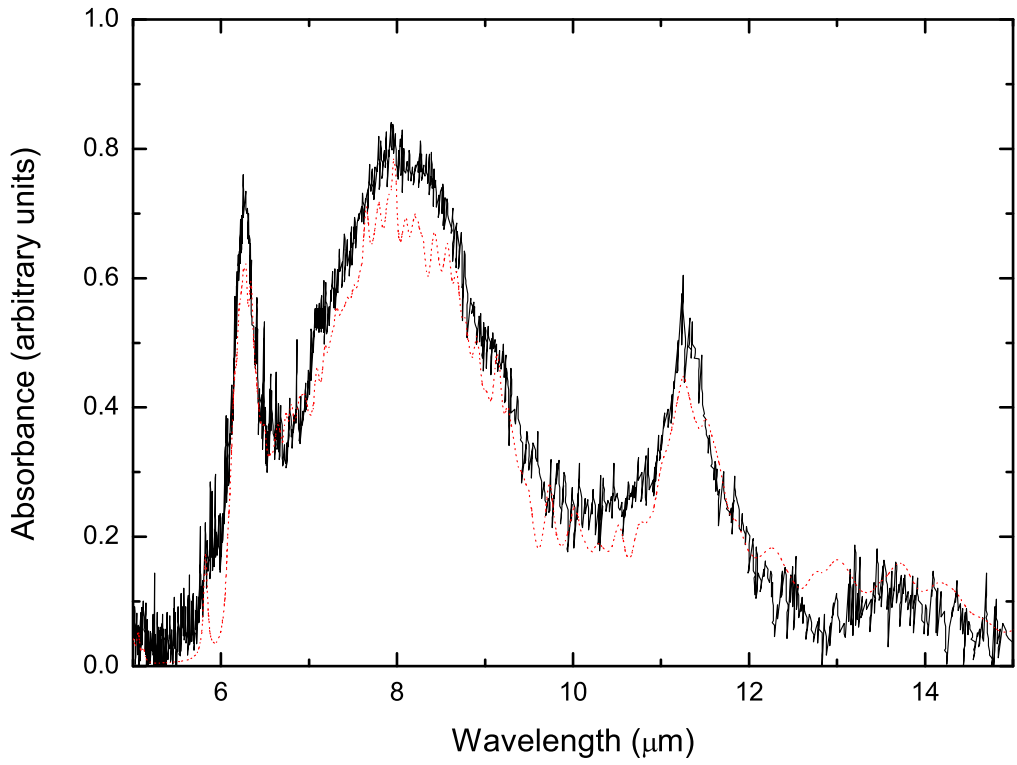, height=12cm}
\caption{Fitting of the absorbance spectra of anthracite coal (solid line, from \citet{guillois1996} using the PAHdb model (dashed line). The value of $\chi^2/dof$ is 0.09.
}
\label{coal}
\end{figure*}

Figure \ref{coal} shows the fit to the absorbance spectrum of anthracite coal \citep{guillois1996} with the PAHdb model. The spectrum shows some resemblance to the HAC spectrum in having broad plateau features.  Again, a mixture of neutral, anion, and cation PAHs can fit the spectrum of coal.  

\begin{figure*}
\epsfig{file=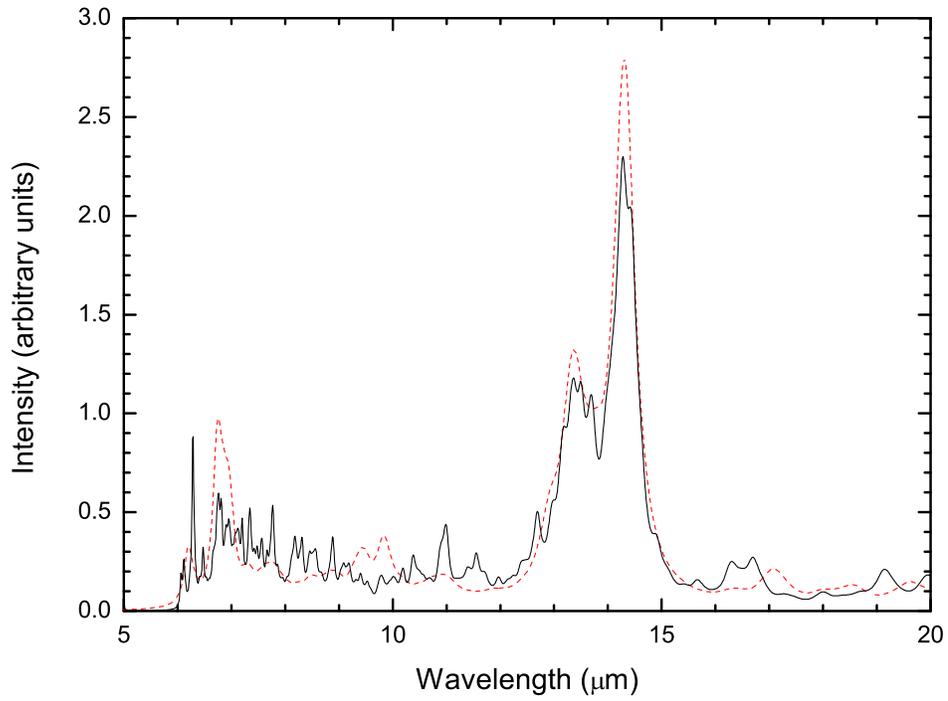, height=11cm}
\caption{Fitting of the spectrum (solid line) of the
$sp^2$/$sp^3$ bonded molecule C$_{55}$H$_{56}$ by the PAHdb model (dashed line).  The value of $\chi^2/dof$ is 0.05.}
\label{maon}
\end{figure*}

\begin{figure*}
\centering
\epsfig{file=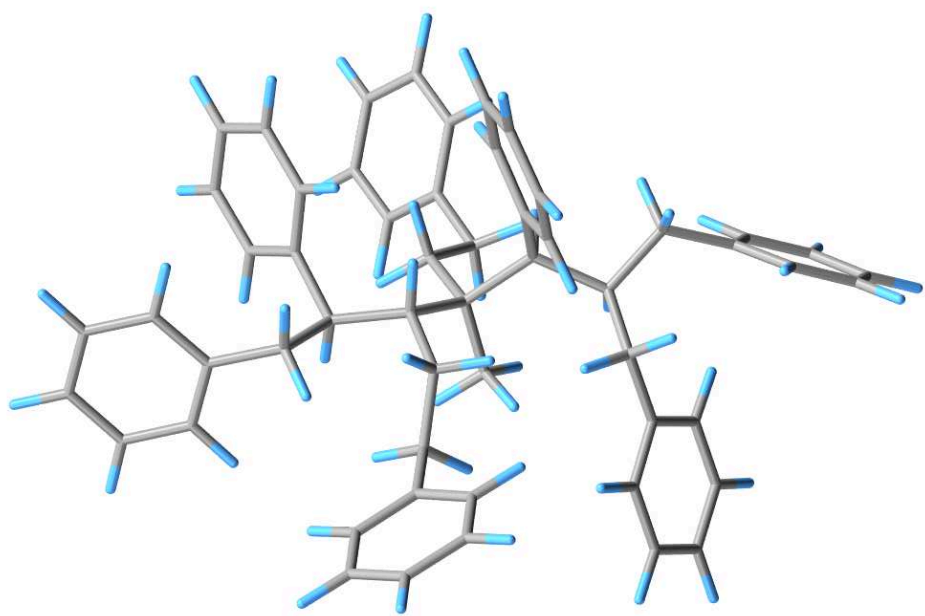, height=8cm}
\caption{Molecular structure of the C$_{55}$H$_{56}$ molecule.
Carbon and hydrogen atoms are represented in grey and blue, respectively.  Figure provided by SeyedAbdolreza Sadjadi.}
\label{c55h56}
\end{figure*}

The PAHdb model can also easily reproduce  the theoretical spectrum of C$_{55}$H$_{56}$ (Figure~\ref{maon}). C$_{55}$H$_{56}$ has a three-dimensional disordered structure with aromatic rings linked by aliphatic chains (Figure \ref{c55h56}), representing a possible partial structure of a MAON particle. The chemical structure of this molecule is completely different from the family of PAH molecules.  Its theoretical spectrum is obtained from the density functional theory following the method described in \citet{sad14}. Although the strong aliphatic feature at 6.9\,$\mu$m cannot be accounted for by PAH mixture, the overall spectrum of C$_{55}$H$_{56}$ can be well reproduced.

\begin{figure*}
\epsfig{file=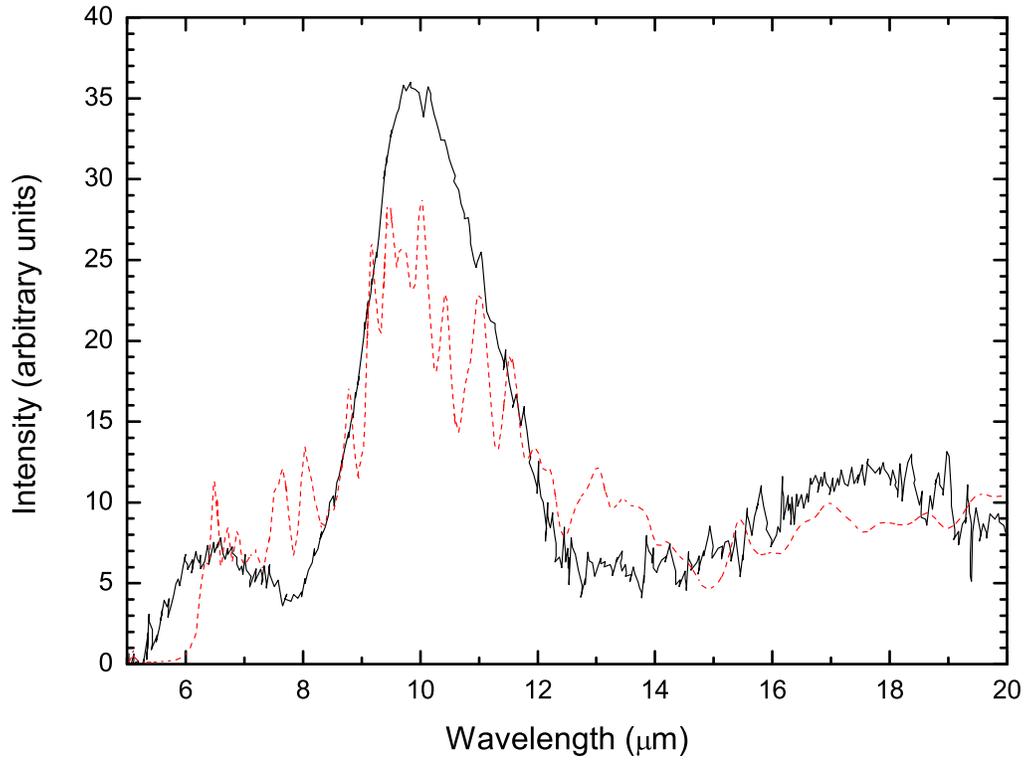, height=12cm}
\caption{PAHdb model fitting (dashed line) of the continuum-subtracted {\it ISO} spectrum (solid line) of the O-rich AGB star V778 Cyg.  The features at 10 and 18 $\mu$m are due to amorphous silicates. The value of $\chi^2/dof$ is 1.56.}
\label{silicate}
\end{figure*}

Figure \ref{silicate} shows the fitting of the emission spectrum from amorphous silicates in the circumstellar envelopes of the asymptotic giant branch (AGB) star V778 Cyg \citep{yama2000}.  Silicate features, in either  emission or absorption, are commonly observed in oxygen-rich AGB stars \citep{kwok1997}.   The spectrum of V778 Cyg shows both the 10 and 18 $\mu$m amorphous silicate features in emission and  the PAHdb model has no problem creating a fit for the spectrum.

\begin{figure*}
\epsfig{file=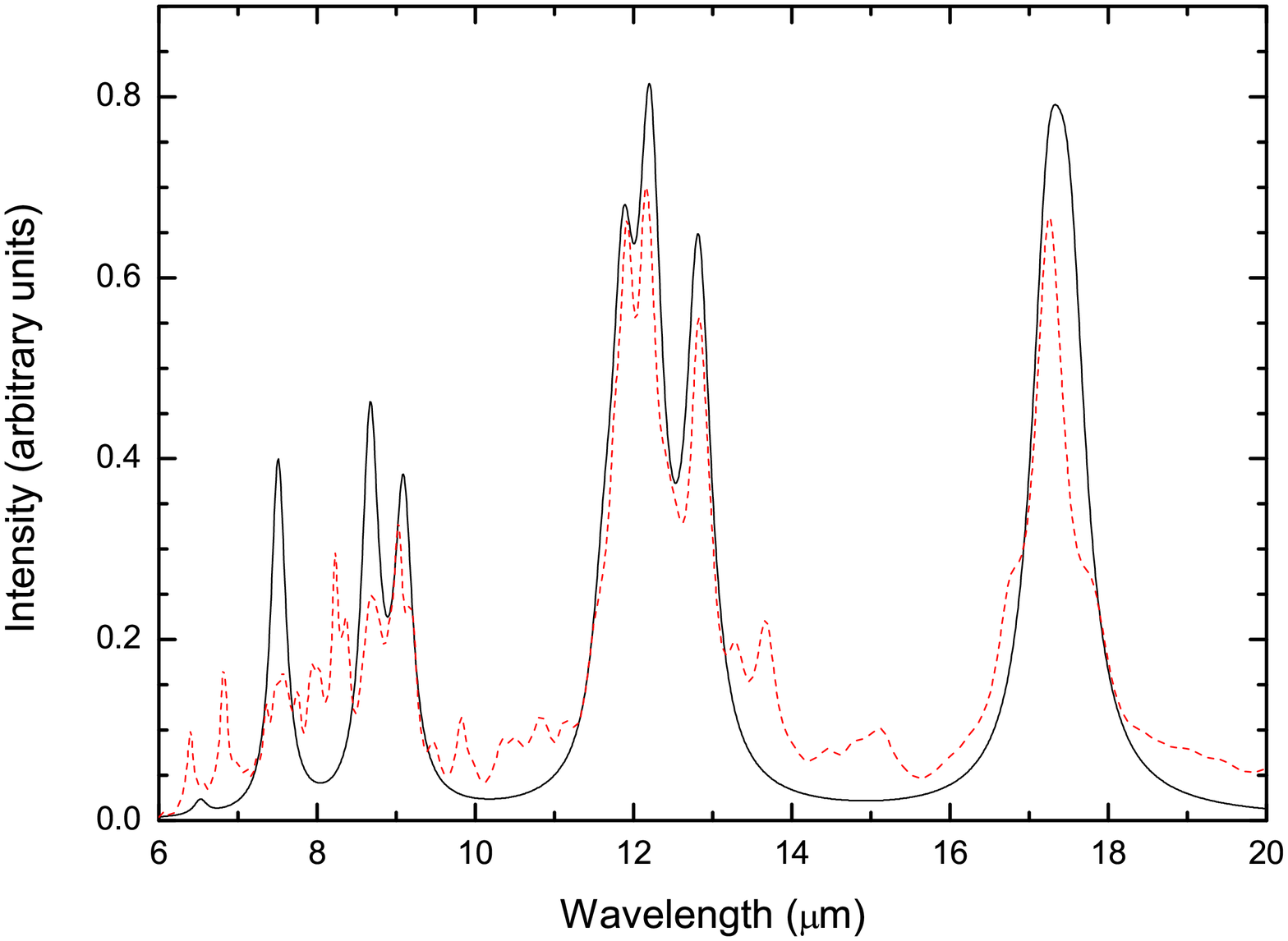, height=12cm}
\caption{PAHdb model fitting (dashed line) of an artificial spectrum (solid line) generated by 10 random features in the 6-20\,$\micron$ range.  The value of $\chi^2/dof$ is 0.04.
}
\label{ran1}
\end{figure*}
\begin{figure*}
\epsfig{file=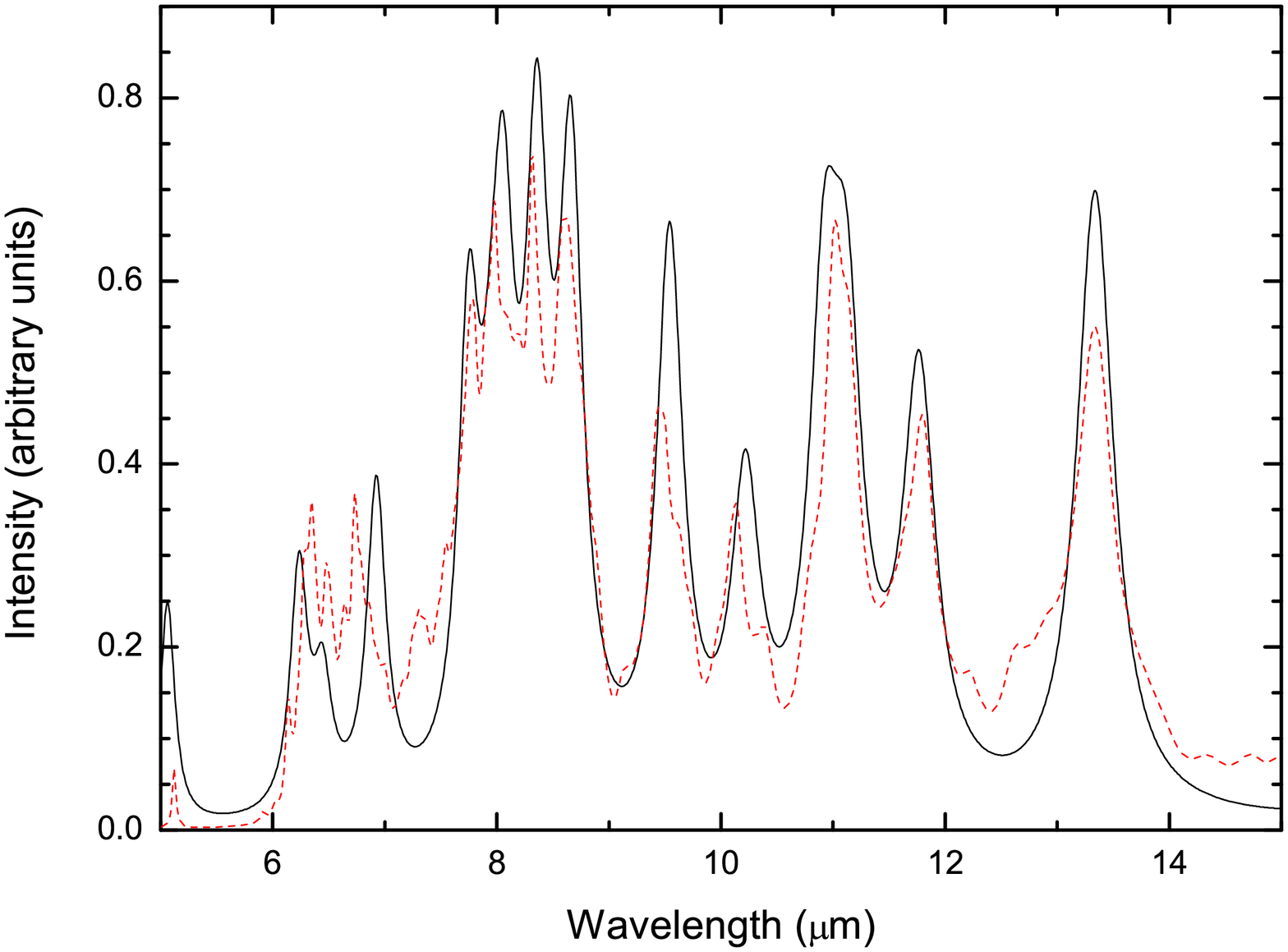, height=12cm}
\caption{PAHdb model fitting (dashed line) of an artificial spectrum (solid line) generated by 20 random features in the 5-15\,$\micron$ range.  The value of $\chi^2/dof$ is 0.11.
}
\label{ran2}
\end{figure*}

We have also  performed a number of exercises where the PAHdb model is used to fit artificially spectra with 10 and 20 randomly generated emission bands.  These features are assumed to have a Drude profile with a fixed fractional width of 0.03, and their wavelengths and intensities
are completely random within the range from 5--15\,$\mu$m.  Examples of two of such fittings are shown in Figures~\ref{ran1} and \ref{ran2}. 
Reasonable matches can be achieved between the artificial spectra and the PAH mixture.  This suggests  that any arbitrary mid-infrared spectra could be successfully simulated using PAH model  through adjusting the weights of the individual PAHs.  In the example of Figure \ref{ran1}, the  strong artificial features at 12 and 17 $\mu$m clearly have nothing to do with aromatic compounds and yet the PAHdb model can fit them well.  In Figure \ref{ran2}, there are multiple strong artificial features and still the PAHdb model is flexible enough to produce a satisfactory fit.  

\begin{figure*}
\epsfig{file=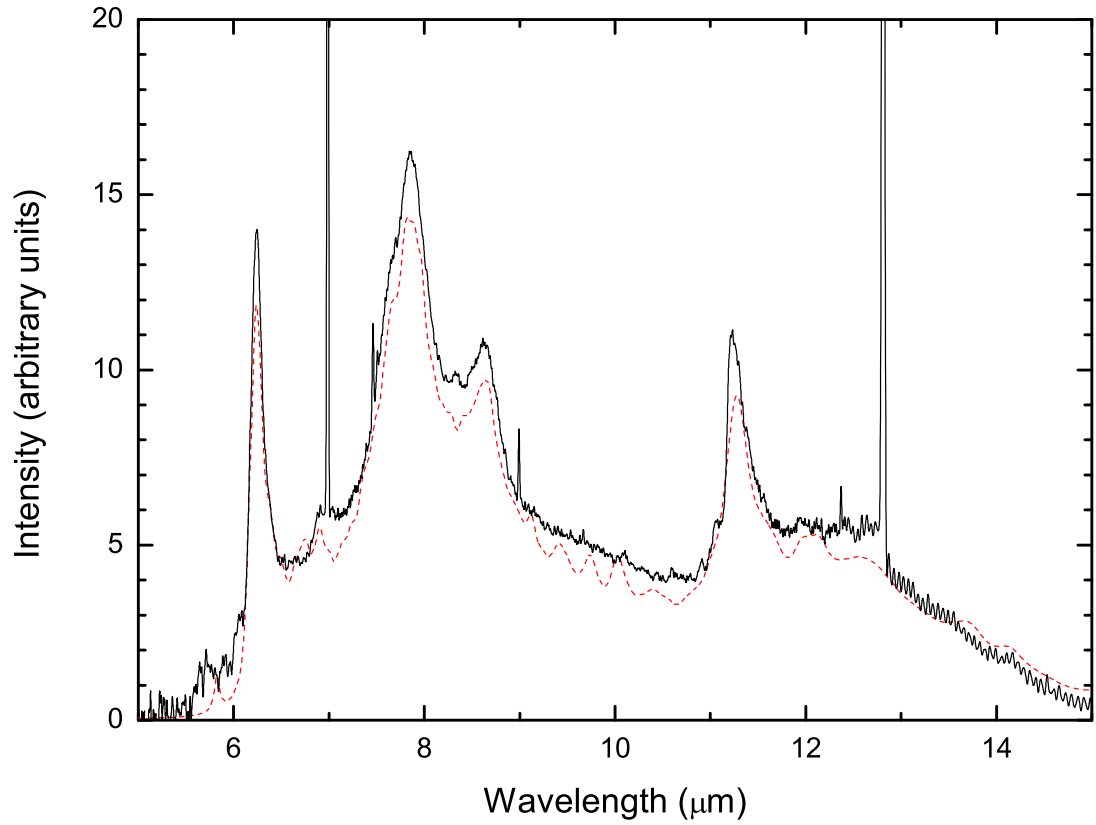, height=13cm}
\caption{Continuum-subtracted {\it ISO} spectrum of BD+30$^\circ$3639 (black curve) fitted by PAHdb model (red curve). The narrow lines in the spectrum of BD+30$^\circ$3639 are atomic lines and are excluded from the fitting.  The value of $\chi^2/dof$ is 0.36.
}
\label{bd30}
\end{figure*}

Although the PAHdb fitting model can produce reasonably good fits to non-PAH spectra as seen in the above examples, it may be argued that the PAHdb model may produce better fits to UIE spectra.  In order to make a quantitative comparison, we have also performed a fitting to the continuum-subtracted   {\it Infrared Space Observatory} ({\it ISO}) spectrum of the planetary nebula BD+30$^\circ$3639 using the PAHdb model (Figure \ref{bd30}).  The continuum component was subtracted using a linear fit to the line-free regions.   The $\chi^2/dof$ value of the fit is 0.36, which is comparable or larger than those for non-PAH spectrum fitting. This suggests that the PAH model does not necessarily generate better match with astronomical spectra than with non-PAH spectra.

\section{Fit of astronomical spectrum by a simple MAON molecule}

\begin{figure*}
\epsfig{file=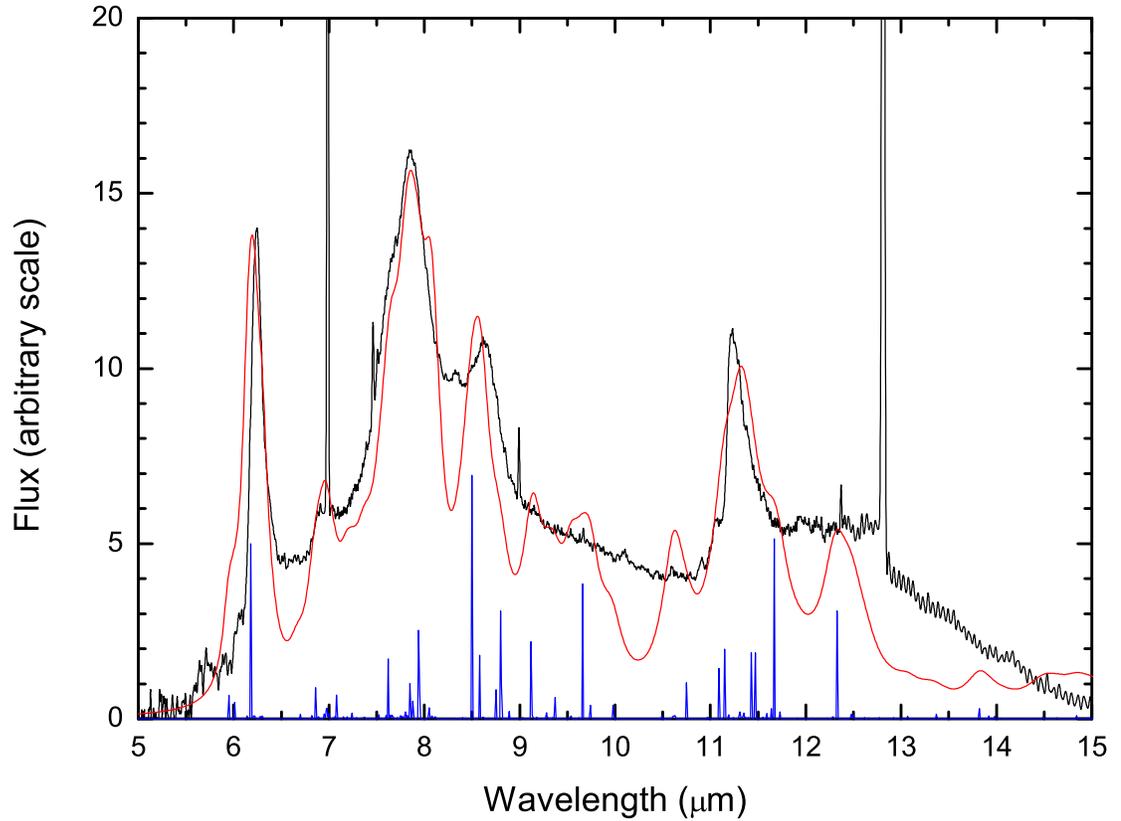, height=13cm}
\caption{Continuum-subtracted {\it ISO} spectrum of BD+30$^\circ$3639 (black curve) fitted by a modified spectrum of C$_{55}$H$_{52}$ (red curve). The relative intensities of theoretical vibrational modes are marked by vertical blue solid lines.  The curve of  C$_{55}$H$_{52}$ is obtained by applying a Drude profile of 500 K to the vibrational lines (for details see \citet{sad14}).  The narrow lines in the spectrum of BD+30$^\circ$3639 are atomic lines and are excluded from the fitting.
}
\label{maonfitting}
\end{figure*}

In this section, we investigate the possibility of fitting  the astronomical spectra of BD+30$^\circ$3639  by a MAON model.     For the model, we chose the theoretical spectrum of C$_{55}$H$_{52}$, which is one of the a series of molecules with mixed aliphatic/aromatic structures calculated by  \citet{sad14}. After applying a Drude profile of 500 K, the spectral regions that contain high density and high intensity modes will have strong bands (Figure \ref{maonfitting}).

The theoretical spectrum of the molecule shows two broad plateau features centered around 7 and 11 $\mu$m, and can be made to fit the observed spectrum by redshifting features with wavelengths larger than 10\,$\mu$m  by 0.4\,$\mu$m.  The strengths of some of the features are also artificially adjusted: the 6.2 $\mu$m feature and the features between 7.5 and 9.4 $\mu$m by a factor of 2.3 and  the 6.9 $\mu$m feature by a factor of 0.5 (Figure \ref{maonfitting}).  Such shifting and adjustment could be the results of anharmonic effect, impurity, geometry, and/or  changes in $sp^2$/$sp^3$ ratio.  Since  C$_{55}$H$_{52}$ is a simple molecule and is only one of many possible MAON structures, we are not claiming that the carrier of UIE bands is this specific molecule.  What this example shows is that instead of involving hundreds of molecules and using equally large number of free parameters, astronomical UIE spectra can be fitted with a single molecule with a few parameters.
This could be a simple coincidence, or the MAON structure has intrinsic properties that are similar to those of the actual carrier of the UIE bands.   
This exercise illustrates that there are other alternatives to fit the UIE bands than relying on mixtures of PAH molecules.

\section{Discussion}

The fact that the PAHdb model can produce fits to HAC, coal, molecules of mixed $sp^2/sp^3$ structure, silicates, and even randomly generated spectra, suggests that the model contains far too many parameters to be considered useful.
One of the objections of \citet{rb14} to the MAON model is that it relies on an empirical decomposition of the observed 
spectra. Here we show that the spectral decomposition based on theoretical PAH spectra does not produce more reliable results than the empirical decomposition.   Through these examinations, we can conclude that a match between astronomical spectra and PAH mixture cannot be taken as an evidence of PAHs as the carrier of UIE.

It is generally acknowledged that no single PAH can reproduce the UIE spectra and a collection of PAH molecules of  various sizes, charge states, and compositions is required to fit astronomical spectra.  So far, there has been no rigorous study on whether the diversity of PAHs is compatible with different astronomical environments.  
Although the UIE bands are found in highly diverse radiation environments, neutral and positive/negative PAH ions are always required to fit the observed spectra.  

As of the to date, there is no experimental support for the synthesis of large groups of diverse PAH molecules under natural conditions.  In contrast, amorphous carbonaceous compounds occur naturally as the result of combustion, or as the result of injection of energy (arc discharge, laser ablation, laser pyrolysis, photolysis, or microwave) into a mixed hydrocarbon gas-phase molecules or graphite \citep{sak87, scott1996, herlin1998,  mennella1999, mennella2003, jager2009, dartois2004, pi08}.

The PAH model has particular difficulty in explaining the plateau features. The plateau features are often explained as the blending of PAH modes, and Lorentzian-like broad wings have to be artificially assumed \citep{ld12}.  In the  MAON model, the plateau features arise naturally from the superposition of many in-plane and out-of-plane bending modes of aliphatic groups as these modes are clustered around 8 and 12 $\mu$m regions \citep{kv01}.   
This is supported by the fact that the UIE and plateau features can be fitted by spectral components decomposed from laboratory spectra of HAC samples  \citep{dh12}.   

If the UIE features are due to PAH molecules, then PAH molecules must be prevalent in the interstellar medium and should be easily detectable in absorption against bright stars.   The observed strengths of the UIE bands would require a  PAH to H$_2$ abundance ratio of $\sim 3\times 10^{-7}$ \citep{tie08}.  
However, no individual PAH molecule has been detected through their electronic, vibrational or rotational transitions \citep{clayton2003, salama2011, gredel2011}.  These failures have been explained by the proposition that the UIE bands are the results of a collection of a large number of diverse PAH molecules, and the abundance of individual PAH molecules is too low to be detected.  Simulations also show that it is impossible to identify individual PAH molecules through their collective spectra in the fitting process of astronomical UIE spectra \citep{rb14}. 

While PAH molecules have narrow vibrational modes occurring at different frequencies, MAONs, in spite of having different chemical compositions, have naturally broad and similar spectral behavior.   
Through the introduction of aliphatic groups, the modes around 8 and 12\,$\mu$m are greatly enriched
due to blended vibrational modes of aliphatic structure and activation of Raman modes.
While a collection of nine small PAHs is not able to reproduce the observed spectrum of BD+30$^\circ$3639 \citep{cs98},   a simple mixed $sp^2/sp^3$ molecule C$_{55}$H$_{52}$ can (Figure~\ref{maonfitting}).  While this in itself may not mean anything, it does suggest that it would be worthwhile to explore the spectral properties of MAONs.  Since MAONs represent a family of amorphous structures, we need to further study how the spectra would vary in response to changes in size, aliphatic/aromatic ratio, and geometry.

\section{Conclusion}

Among the candidate carriers of UIE, the PAH model has attracted the most attention, in part because of its aesthetic appeal.  PAH molecules are small and simple gas-phase molecules and their presence in space is easier to accept than complex organics.  But we now know that complex organics of abiological origin are prevalent in the Solar System.  Because the PAH hypothesis does not specify any single molecule but refers to a group of  large number of diverse molecules, it does not produce concrete predictions for testing or falsification.  Although a mixture of many PAHs can reproduce the UIE spectra, we have shown that such a mixture is also able to produce non-PAH spectra, even randomly generated spectra. Thus the observational foundation of the PAH model is not solid.  

Since the UIE phenomenon is seen throughout the Universe, even during its early epochs, a correct identification of the carrier is of great importance.  Due to the strengths of the UIE bands, the carrier represents a major reservoir of carbon. Whether the carriers are a collection of free-flying gas-phase PAH molecules or complex organic solid particles has significant different implications on our understanding of cosmic chemical synthesis, energy exchange between stars and the interstellar medium,  and galactic chemical enrichment.  The PAH hypothesis has been entrenched in models of the ISM that PAH molecules are believed to be dominant factors in the photoelectric heating of interstellar gas and a determining factor in the ionization balance inside molecular clouds \citep{pah}.  In this paper, we try to show that the situation is not as simple as it has been represented in the literature.  A correct interpretation of the UIE spectrum requires further studies, both computationally and experimentally, of the vibrational properties of amorphous carbons of mixed aromatic/aliphatic structures.    

{\flushleft \bf Acknowledgment~}

We thank SeyedAbdolreza Sadjadi and Renaud Papoular for helpful discussions.  The fitting models were done using the NASA Ames Research Center PAH IR Spectroscopic Database and software and we thank Christiaan Boersma  for making these resources publicly available.  
This work was partially supported by the Research Grants Council of the Hong Kong Special Administrative Region, China (project no. HKU 7031/10P).

\end{CJK*}


\begin{thebibliography}{}

\bibitem[Allamandola et al.(1999)]{ah99} Allamandola, L.~J., 
Hudgins, D.~M., \& Sandford, S.~A.\ 1999, \apjl, 511, L115 

\bibitem[Allamandola et al.(1985)]{al85} Allamandola, L.~J.,
Tielens, A.~G.~G.~M., \& Barker, J.~R.\ 1985, \apjl, 290, L25

\bibitem[Allamandola et al.(1989)]{al89} Allamandola, L.~J.,
Tielens, A.~G.~G.~M., \& Barker, J.~R.\ 1989, \apjs, 71, 733

\bibitem[Bauschlicher et al.(2010)]{bb10} 
Bauschlicher, C.~W., Jr., Boersma, C., Ricca, A., et al.\ 2010, \apjs, 189, 341

\bibitem[Bern{\'e} et al.(2007)]{ber07} Bern{\'e}, O., Joblin, C., Deville, Y., et al.\ 2007, \aap, 469, 575 

\bibitem[Boersma et al.(2013)]{bb13} 
Boersma, C., Bregman, J.~D., \& Allamandola, L.~J.\ 2013, \apj, 769, 117 

\bibitem[Boersma et al.(2014)]{bb14} Boersma, C., 
Bauschlicher, C.~W., Jr., Ricca, A., et al.\ 2014, \apjs, 211, 8 


\bibitem[Cami(2011)]{ca11} Cami, J. 2011, in EAS Publications Series, Vol. 46, PAHs and the Universe, ed.
C. Joblin \& A. G. G. M. Tielens (Cambridge: Cambridge Univ. Press), 117

\bibitem[Cataldo, \& Keheyan(2003)]{cataldo2003} 
Cataldo, F., \& Keheyan, Y. 2003, International Journal of Astrobiology, 2, 41

\bibitem[Chiar et al.(2000)]{ct00} Chiar, J.~E., Tielens, 
A.~G.~G.~M., Whittet, D.~C.~B., et al.\ 2000, \apj, 537, 749 

\bibitem[Clayton et al.(2003)]{clayton2003} 
Clayton, G. C., et al. 2003, \apj, 592, 947

\bibitem[Colangeli et al.(1997)]{col1997}
Colangeli, L., Bussoletti, E., Pestellini, C. C., Mennella, V., Palomba, E., Palumbo, P., \& Rotundi, A. 1997, Advances in Space Research, 20, 1617

\bibitem[Cook \& Saykally(1998)]{cs98} 
Cook, D.~J., \& Saykally, R.~J.\ 1998, \apj, 493, 793 

\bibitem[Dartois et al.(2004)]{dartois2004}
Dartois, E., Mu\~noz Caro, G. M., Deboffle, D., \& d'Hendecourt, L. 2004, \aap, 423, L33

\bibitem[Dischler et al.(1983)]{dis83}
Dischler, B., Bubenzer, A.,\& Koidl, P. 1983, Sol. Sta. Com., 48, 105

\bibitem[Duley 
\& Hu(2012)]{dh12} Duley, W.~W., \& Hu, A.\ 2012, \apj, 761, 115 

\bibitem[Duley \& Williams(1981)]{dw81} 
Duley, W.~W., \& Williams, D.~A.\ 1981, \mnras, 196, 269

\bibitem[Duley, \& Williams(1983)]{duley1983} 
Duley, W. W., \& Williams, D. A. 1983, \mnras, 205, 67P

\bibitem[Gredel et al.(2011)]{gredel2011}
Gredel, R., Carpentier, Y., Rouill\'e, G., Steglich, M., Huisken, F., \& Henning, T. 2011, \aap, 530, 26

\bibitem[Guillois et al.(1996)]{guillois1996}
 Guillois, O., Nenner, I., Papoular, R., \& Reynaud, C. 1996, \apj, 464, 810

\bibitem[Herlin et al.(1998)]{herlin1998} 
Herlin, N., Bohn, I., Reynaud, C., Cauchetier, M., Galvez, A., \& Rouzaud, J.-N. 1998, \aap, 330, 1127

\bibitem[J\"ager et al.(2009)]{jager2009} 
J\"ager, C., Huisken, F., Mutschke, H., Jansa, I. L., \& Henning, T. H. 2009, \apj, 696, 706


\bibitem[Joblin \& Tielens(2011)]{pah}
Joblin, C., \& Tielens, A.G.G.M. 2011, PAHs and the Universe, EDP Sciences

\bibitem[Jourdain de Muizon et 
al.(1990)]{jd90} Jourdain de Muizon, M., D'Hendecourt, L.~B., \& Geballe, T.~R.\ 1990, \aap, 235, 367 

\bibitem[Kwok, Volk, \& Bidelman(1997)]{kwok1997} 
Kwok, S., Volk, K., \& Bidelman, W. P. 1997, \apjs, 112, 557

\bibitem[Kwok et al.(2001)]{kv01} 
Kwok, S., Volk, K., \& Bernath, P.\ 2001, \apjl, 554, L87

\bibitem[Kwok et 
al.(1999)]{kv99} Kwok, S., Volk, K., \& Hrivnak, B.~J.\ 1999, \aap, 350, L35 

\bibitem[Kwok \& Zhang(2011)]{kz11} Kwok, S., \& Zhang, Y.\ 2011, \nat, 479, 80 

\bibitem[Kwok \& Zhang(2013)]{kz13} 
Kwok, S., \& Zhang, Y.\ 2013, \apj, 771, 5 


\bibitem[L\'{e}ger \& Puget(1984)]{lp84} 
L\'{e}ger, A., \& Puget, J.~L.\ 1984, \aap, 137, L5

\bibitem[Li \& Draine(2012)]{ld12} 
Li, A., \& Draine, B.~T.\ 2012, \apjl, 760, L35 

\bibitem[Mennella et al.(1999)]{mennella1999} 
Mennella, V., Brucato, J. R., Colangeli, L., \& Palumbo, P. 1999, \apj, 524, L71

\bibitem[Mennella et al.(2003)]{mennella2003}
Mennella, V., Baratta, G. A., Esposito, A., Ferini, G., \& Pendleton, Y. J. 2003, \apj, 587, 727

\bibitem[Papoular et al.(1989)]{papoular1989} 
Papoular, R., Conrad, J., Giuliano, M., Kister, J., \& Mille, G. 1989, \aap, 217, 204

\bibitem[Peeters(2013)]{pe13} Peeters, E.\ 2013,  in IAU Symp. 297,
The Diffuse Interstellar Bands, eds. J. Cami \& N. L. J. Cox
(Cambridge: Cambridge Univ. Press), 187

\bibitem[Pilleri et al.(2012)]{pil12} Pilleri, P., Montillaud, J., Bern{\'e}, O., \& Joblin, C.\ 2012, \aap, 542, 69 

\bibitem[Pino et al.(2008)]{pi08} 
Pino, T., Dartois, E., Cao, A.-T., et al.\ 2008, \aap, 490, 665 

\bibitem[Puget \& L\'{e}ger(1989)]{pl89} Puget, J.~L., \& L\'{e}ger, A.\ 1989, \araa, 27, 161


\bibitem[Rosenberg et 
al.(2011)]{rb11} Rosenberg, M.~J.~F., Bern{\'e}, O., Boersma, C., Allamandola, L.~J., \& Tielens, A.~G.~G.~M.\ 2011, \aap, 532, 128 

\bibitem[Rosenberg et 
al.(2014)]{rb14} Rosenberg, M.~J.~F., Bern{\'e}, O., \& Boersma, C.\ 2014, \aap, 566, L4 

\bibitem[Sadjadi et al.(2014)]{sad14} 
Sadjadi, S., Zhang, Y., \& Kwok, S.\ 2014, \apj, submitted

\bibitem[Sakata et al.(1987)]{sak87}
Sakata, A., Wada, S., Onaka, T., \& Tokunaga, A. T. 1987, \apj, 320, L63

\bibitem[Salama et al.(2011)]{salama2011}
Salama, F., Galazutdinov, G. A., Krelowski, J., Biennier, L., Beletsky, Y., \& Song, I.-O. 2011, \apj, 728, 154

\bibitem[Scott \& Duley(1996)]{scott1996} Scott, A., \& Duley, W. W. 1996, \apj, 472, L123

\bibitem[Sellgren, Uchida, \& Werner(2007)]{sellgren2007} 
Sellgren, K., Uchida, K. I., \& Werner, M. W. 2007, \apj, 659, 1338


\bibitem[Sturm et al.(2000)]{sturm2000} 
Sturm, E., et al. 2000, \aap, 358, 481

\bibitem[Tielens(2008)]{tie08} Tielens, A.~G.~G.~M.\ 2008, \araa, 46, 289

\bibitem[Yamamura et al.(2000)]{yama2000} 
Yamamura, I., Dominik, C., de Jong, T., Waters, L.~B.~F.~M., \& Molster, F.~J.\ 2000, \aap, 363, 629 

\bibitem[Yang et al.(2013)]{yg13} Yang, X.~J., Glaser, R., 
Li, A., \& Zhong, J.~X.\ 2013, \apj, 776, 110 

\end{thebibliography}
\end{document}